\documentclass{elsarticle}

\usepackage{graphicx}
\usepackage{bm}
\usepackage{amssymb}
\usepackage{amsmath}
\usepackage{amsfonts}

\usepackage[utf8]{inputenc}

\begin{document}

\begin{frontmatter}

\title{Nonlinear internal gravity waves in the atmosphere: Rogue waves, breathers and dark solitons}

\author[mymainaddress,mysecondaryaddress]{Volodymyr M. Lashkin \corref{mycorrespondingauthor}}
\cortext[mycorrespondingauthor]{Corresponding author}
\ead{vlashkin62@gmail.com}

\author[mysecondaryaddress]{Oleg K. Cheremnykh}

\address[mymainaddress]{Institute for Nuclear Research, Pr. Nauki 47, Kyiv 03028,
Ukraine}
\address[mysecondaryaddress]{Space Research Institute, Pr. Glushkova 40
k.4/1, Kyiv 03187,  Ukraine}

\begin{abstract}
We study nonlinear internal gravity waves (IGWs) in the
atmosphere. The reductive perturbation method is used to derive a
system of two-dimensional nonlinear equations for the envelope of
velocity stream function and the mean flow. In the one-dimensional
case, we obtain a nonlinear Schr\"{o}dinger (NLS) equation
corresponding to both horizontal and vertical propagation of IGWs.
Depending on the characteristic wavelengths, the NLS equation is
focusing or defocusing. In the focusing case, non-stationary
solutions in the form of the Peregrine soliton, the Akhmediev
breather and the Kuznetsov-Ma breather are considered as potential
candidates for the modeling of rogue waves in the atmosphere. In
the defocusing case, stationary nonlinear IGWs are considered in
the form of nonlinear periodic waves and dark solitons.
\end{abstract}

\begin{keyword}
internal gravity waves  \sep atmosphere \sep rogue waves \sep
breather \sep dark soliton
\end{keyword}
\end{frontmatter}

\section{Introduction}

Internal gravity waves (IGWs) in the atmosphere of the Earth, in
the solar atmosphere, as well as in planetary atmospheres
constitute the most intense part of the spectrum of
acoustic-gravity waves and have been the subject of a large number
of experimental and theoretical studies for many years
\cite{Hines1960,Tolstoy1967,Liu1974,Beer1974,Gossard1975}. The
IGWs are low frequency disturbances associated with the density
and velocity perturbations of the atmospheric fluid in the
presence of the equilibrium pressure gradient that is maintained
by the gravity force. These waves play a significant role in the
formation of atmospheric convection and turbulence and have an
essential influence both on a dynamics of the atmosphere and
coupling of the upper atmosphere with ionosphere. The study of
IGWs is also motivated by the need to obtain accurate predictions
of atmospheric dynamics under various meteorological conditions.

The linear theory of IGWs  has been developed in great detail
(see, e.g., \cite{Sutherland2015}, reviews
\cite{Francis1975,Fritts2003} and references therein). Such
effects as the modulation instability of IGWs leading to the
emergence of zonal flows \cite{Horton-zonal2008}, the influence of
the Coriolis and Amp\'{e}re forces (for the case of an ionized
atmosphere) \cite{Kaladze2008}, existence of evanescent
acoustic-gravity waves with a continuous spectrum
\cite{Cheremnykh2021}, and the presence of a random temperature
profile resulting in the threshold instability of IGWs
\cite{Lashkin2023} were studied.

In many cases, however, it is not possible to confine ourselves to
considering only linear IGWs. The dynamics of the atmosphere is
governed by the totality of all motions, taking into account their
nonlinear interaction. In particular, in the Earth's atmosphere,
the amplitudes of IGWs grow exponentially with increasing
altitude. Finite amplitude IGWs in the atmosphere have been
considered in a fairly large number of works. The resonant and
nonresonant interactions between gravity waves and vortical modes
in the atmosphere were investigated in \cite{Dong1988,Fritts1992}.
The nonlinear ionospheric response to IGWs was studied in
\cite{Huang1991} and distortions in the waveform of ionospheric
disturbances caused by nonlinear effects are predicted. In
\cite{Huang1992}, the interaction of atmospheric gravity waves
with ion-acoustic waves in the $F$ region of the ionosphere was
studied and a coupled pair of Korteweg-de Vries equations was
derived. It was shown that nonlinear atmospheric gravitational
solitary waves can be excited as a result of ion-neutral
collisions. In \cite{Nekrasov1994}, a nonlinear saturation of
atmospheric gravity waves was considered and it was shown that the
amplitude of the vertical velocity perturbation of IGW which would
exponentially grow with altitude in the linear approximation was
restricted by a nonlinear stabilization. The stabilization of the
collapse (breaking) of the nonlinear IGW in an inhomogeneous
atmosphere due to the effects of viscosity was discussed in
\cite{Nekrasov2005}. Intensive numerical modeling of the dynamics
of atmospheric IGWs in the framework of nonlinear fluid equations
was carried out in
\cite{Gavrilov2005,Huang2014,Fritts2015,Snively2017,Fritts2019}.
Simplified two-dimensional and three-dimensional nonlinear
equations for describing the dynamics of IGWs in the atmosphere
were obtained by Stenflo
\cite{Stenflo1987,Stenflo1990,Stenflo2009}. Based on these
equations, vortex-like coherent nonlinear structures of IGWs were
studied. Two-dimensional dipole vortices in the form of a
cyclone-anticyclone pair, analogous to Larichev-Reznik solitons
(modons), were found analytically
\cite{Stenflo1987,Stenflo2009,Shukla1998,Fedun2013}. Solutions in
the form of tripole vortices and vortex chains of IGWs were
obtained in \cite{Jovanovic2001,Jovanovic2002}. Nonlinear IGWs
were also considered in \cite{Fedun2016,Fedun2021}, where,
neglecting dispersion, the so-called dust devils (rotating columns
of rising dust) were studied. Recently, the two-dimensional
Stenflo equations have been generalized to the case of a weakly
ionized ionosphere, taking into account the Amp\'{e}re force,
transverse (Pedersen) and Hall conductivities, and solutions in
the form of dipole vortices have also been found
\cite{Misra2022IEEE,Misra2022AdvSpace}.

One of the remarkable and intriguing phenomena discovered in
recent years in fluid physics is the possibility of rogue waves
(also known as "freak"  waves or "killer"  waves). The rogue wave
is a short-lived high-amplitude wave that suddenly appears against
a constant background and then disappears. Rogue waves are now
recognized as proper intrinsically nonlinear structures (beyond an
initial attempt to identify them as superposed linear modes).
First discovered in the ocean \cite{Dysthe2008,Pelinovski2009},
these waves were subsequently experimentally discovered and then
theoretically studied in optics
\cite{Solli2007,Frisquet2016,Baronio2018}, superfluid helium
\cite{Ganshin2008}, Bose-Einstein condensates \cite{Bludov2009},
plasmas \cite{Shukla2011,Bailung2011}, molecular systems during
chemical reaction \cite{Tlidi2016}, and even finance
\cite{Yan2011}. However, as far as we know, no theoretical studies
of rogue waves in the atmosphere have been reported yet, with the
exception of a short report by Stenflo and Marklund
\cite{Marklund2010}, where it is simply indicated that the
description of rogue waves in the ocean by the nonlinear
Schr\"{o}dinger (NLS) equation is very similar to the description
of atmospheric disturbances and it is noted that the study of
these nonlinear wave structures in the atmosphere is of undoubted
interest.

In this paper, we consider the Stenflo equations for atmospheric
IGWs in the envelope approximation and, using the reductive
perturbation method, derive a system of two-dimensional nonlinear
equations for the velocity stream function and the mean flow. In
the one-dimensional case, we obtain the NLS equation for the
envelope corresponding to both horizontal and vertical propagation
of IGWs. Depending on the ratio of horizontal and vertical
wavelengths, this equation can have both focusing (the signs of
the dispersion and nonlinear terms are the same) and defocusing
type. In the focusing case, non-stationary solutions in the form
of the Peregrine soliton (rogue wave), the Akhmediev breather and
the Kuznetsov-Ma breather are considered as potential candidates
for the modeling of rogue waves in the atmosphere. In the
defocusing case, stationary nonlinear IGWs are considered in the
form of nonlinear periodic waves and dark solitons.

The paper is organized as follows. In Section \ref{Sec2} the
Stenflo equations are presented and commented. Reductive
perturbation analysis is given in Section \ref{Sec3}. In Section
\ref{Sec4} we derive focusing and defocusing NLS equations.
Solutions in the form of the breathers and Peregrine soliton are
presented in Section \ref{Sec5}, and the nonlinear periodic waves
and dark solitons are considered in Section \ref{Sec6}. The
conclusion is made in Section \ref{Sec7}.

\section{\label {Sec2} Model equations}

Nonlinear Stenflo equations \cite{Stenflo1987,Stenflo2009}
governing the dynamics of atmospheric IGWs in the two-dimensional
version have the form
\begin{gather}
\frac{\partial}{\partial
t}\left(\Delta\psi-\frac{1}{4H^{2}}\psi\right)+\{\psi,\Delta\psi\}+\frac{\partial
\chi}{\partial x}=0, \label{main1}
\\
\frac{\partial \chi}{\partial
t}+\{\psi,\chi\}-\omega_{g}^{2}\frac{\partial \psi}{\partial x}=0,
\label{main2}
\end{gather}
where $\Delta=\partial^{2}/\partial x^{2}+\partial^{2}/\partial
z^{2}$ is the two-dimensional Laplasian, and the Poisson bracket
(the Jacobian) $\{f,g\}$ defined by
\begin{equation}
\{f,g\}=\frac{\partial f}{\partial x}\frac{\partial g}{\partial z}
-\frac{\partial f}{\partial z}\frac{\partial g}{\partial x}.
\end{equation}
Here, $\psi(x,z)$ is the velocity stream function, $\chi (x,z)$ is
the normalized density perturbation, $H$ is the density scale
height (reduced atmospheric height), $\omega_{g}=(g/H)^{1/2}$ is
the Brunt-V\"{a}is\"{a}l\"{a} or buoyancy frequency, $g$ is the
free fall acceleration. Equations (\ref{main1}) and (\ref{main2})
depend only on two Cartesian coordinates $x$ and $z$, where the
$z$ axis is directed upward against the gravitational acceleration
$\mathbf{g}=-g\hat{\mathbf{z}}$, where $\hat{\mathbf{z}}$ is the
unit vector along the $z$ direction and the $x$ axis lies in a
plane perpendicular to the $z$ axis. They do not take into account
the curvature of the planet and the rotation of the atmosphere.
Therefore, the atmosphere in the $(x, y)$ plane is considered
isotropic and the dependence on the coordinate $y$ can be
eliminated by the corresponding rotation of the coordinate system
around the axis so that the $x$ axis is directed along the
horizontal component of the fluid velocity, so that
$v_{x}=\partial\psi/\partial z$ and $v_{z}=-\partial\psi/\partial
x$.

In the linear approximation, taking $\psi\sim\exp
(i\mathbf{k}\cdot\mathbf{x}-i\omega t)$ and $\chi\sim\exp
(i\mathbf{k}\cdot\mathbf{x}-i\omega t)$, where $\mathbf{x}=(x,z)$,
$\omega$ and $\mathbf{k}=(k_{x},k_{z})$ are the frequency and wave
number respectively, Eqs. (\ref{main1}) and (\ref{main2}) yield
the dispersion relation of the gravity waves
\begin{equation}
\label{dispers}
\omega^{2}=\frac{\omega_{g}^{2}k_{x}^{2}}{k^{2}+1/(4H^{2})},
\end{equation}
where $k^{2}=k_{x}^{2}+k_{z}^{2}$. In Eqs. (\ref{main1}) and
(\ref{main2}), the Coriolis force is neglected, and in the IGWs
dynamics is valid for $\omega\gg\Omega_{0}$, where $\Omega_{0}$ is
the angular rotation velocity of the planet. Thus, we exclude from
consideration the case of very small horizontal wave numbers
$k_{x}\ll \omega_{g}\Omega_{0}/g$. We also consider altitudes at
which the Amp\'{e}re force can be neglected, and where the effect
of the geomagnetic field is of the same order as the effect due to
the Coriolis force \cite{Kaladze2008}. In addition, the
Brunt-V\"{a}is\"{a}l\"{a} frequency is assumed to be independent
of the vertical coordinate $z$, that is, further we consider an
isothermal atmosphere. For the Earth's atmosphere, in particular,
this corresponds to altitudes  $\gtrsim 200$ km. Then, the lower
limit for wavelengths (due to the dissipation of short-wave
harmonics) for IGWs is about $\sim 10$ km at altitudes
 $\sim 200$-$300$ km while typical characteristic values are
hundreds of kilometers.

\section{\label {Sec3} Reductive perturbation analysis}

To investigate the nonlinear behavior of the IGWs, we use
reductive perturbation method (sometimes also called the
multiscale expansion method) \cite{Dodd1982} which is often used
in the theory of nonlinear waves. This method usually leads to
asymptotic evolution equations, sometimes more adequate to the
given problem. Following this technique, we expand the space and
time variables as
$\mathbf{x}=\mathbf{x}+\varepsilon\mathbf{X}+\dots$ and
$t=t+\varepsilon T+\varepsilon^{2}\tau+\dots$ respectively, where
$\mathbf{X}=(X,Z)$, and $\varepsilon$ is the small dimensionless
parameter scaling the weakness of dispersion and nonlinearity. As
will be shown later, to obtain a non-trivial evolution, it
suffices to restrict ourselves to expanding the time variable up
to the second order and the space variable up to the first order
in $\varepsilon$. Thus, we have
\begin{equation}
\label{exp-x-t} \frac{\partial}{\partial \mathbf{x}}\rightarrow
\frac{\partial}{\partial \mathbf{x}}+\varepsilon
\frac{\partial}{\partial \mathbf{X}}, \quad
\frac{\partial}{\partial t}\rightarrow \frac{\partial}{\partial
t}+\varepsilon \frac{\partial}{\partial T}+\varepsilon^{2}
\frac{\partial}{\partial \tau}.
\end{equation}
We then expand the fields $\psi$ and $\chi$ in powers in
$\varepsilon$ as
\begin{gather}
\psi=\varepsilon\psi^{(1)}+\varepsilon^{2}\psi^{(2)}+\varepsilon^{3}\psi^{(3)}+\dots,
\label{exp-psi}
\\
\chi=\varepsilon\chi^{(1)}+\varepsilon^{2}\chi^{(2)}+\varepsilon^{3}\chi^{(3)}+\dots,
\label{exp-phi}
\end{gather}
where $\psi^{(1)}=\tilde{\psi}^{(1)}+\bar{\psi}$,
$\chi^{(1)}=\tilde{\chi}^{(1)}+\bar{\chi}$,
\begin{gather}
\tilde{\psi}^{(1)}=\Psi
(\mathbf{X},T,\tau)\mathrm{e}^{i\mathbf{k}\cdot\mathbf{x}-i\omega
t}+\mathrm{c. c.}, \label{Psi}
\\
\tilde{\chi}^{(1)}=\Phi
(\mathbf{X},T,\tau)\mathrm{e}^{i\mathbf{k}\cdot\mathbf{x}-i\omega
t}+\mathrm{c. c.}. \label{Phi}
\end{gather}
Secondary mean flows $\bar{\psi}$ and $\bar{\chi}$ depend only on
slow variables $\mathbf{X}$, $T$ and $\tau$. Our goal is to obtain
nonlinear evolution equation for the envelope $\Psi$. Acting by
the operator $\partial/\partial t$ on Eq. (\ref{main1}), and then
using Eq. (\ref{main2})), one can obtain
\begin{equation}
\label{main3} \mathcal{L}\psi=\mathcal{N},
\end{equation}
where
\begin{gather}
\mathcal{L}=\frac{\partial}{\partial
t^{2}}\left(\Delta-\frac{1}{4H^{2}}\right)+\omega_{g}^{2}\frac{\partial^{2}}{\partial
x^{2}}, \label{L}
\\
\mathcal{N}=\frac{\partial}{\partial
x}\{\psi,\chi\}-\frac{\partial}{\partial t}\{\psi,\Delta\psi\},
\label{N}
\end{gather}
and the linear part of Eq. (\ref{main3}) contains only $\psi$. The
operator
$\mathcal{L}(\partial_{t}+\varepsilon\partial_{T}+\varepsilon^{2}\partial_{\tau},
\partial_{\mathbf{x}}+\varepsilon\partial_{\mathbf{X}})$ can be
expanded in terms of $\varepsilon$,
$\mathcal{L}=\mathcal{L}_{0}+\epsilon\mathcal{L}_{1}+\varepsilon^{2}\mathcal{L}_{2}$.
Then substituting Eqs. (\ref{exp-x-t}), (\ref{exp-psi}) and
(\ref{exp-phi}) into Eq. (\ref{main3})  and keeping terms up to
first order in $\varepsilon$, we get
\begin{equation}
\label{first-order} \mathcal{L}_{0}\tilde{\psi}^{(1)}=0, \quad
\mathcal{L}_{0}\tilde{\chi}^{(1)}=0,
\end{equation}
that is
$\mathcal{L}_{0}(\omega,\mathbf{k})=\omega^{2}(k^{2}+1/4H^{2})-\omega_{g}^{2}k_{x}^{2}=0$
gives the dispersion relation (\ref{dispers}). In the next order
$O(\varepsilon^{2})$ we have
\begin{equation}
\label{second-order}
\mathcal{L}_{0}\tilde{\psi}^{(2)}=\mathcal{L}_{1}\tilde{\psi}^{(1)},
\end{equation}
or
\begin{equation}
\label{second-order1} \mathcal{L}_{0}\tilde{\psi}^{(2)}=
\left(\frac{\partial\mathcal{L}_{0}}{\partial\omega}\frac{\partial}{\partial
T}+\frac{\partial\mathcal{L}_{0}}{\partial\mathbf{k}}\frac{\partial}{\partial\mathbf{X}}\right)
\tilde{\psi}^{(1)}.
\end{equation}
Similar equations hold for $\chi$. Note that despite the quadratic
nature of the nonlinearity in Eq. (\ref{main3}), the right hand
sides of Eq. (\ref{second-order}) in the $\varepsilon^{2}$ order
do not contain nonlinear terms. This is due to the specific type
of nonlinearity in Eq. (\ref{main3}) in the form of the Poisson
bracket, when the corresponding nonlinear terms disappear
identically. As usual \cite{Dodd1982}, the $\varepsilon^{2}$ order
secular terms, that is the right hand side of Eq.
(\ref{second-order1}), represent the group motion of the envelope
and can be eliminated by transforming to a frame moving with the
group velocity
\begin{equation}
\label{group-vel}
\mathbf{v}_{g}=\frac{\partial\mathcal{L}_{0}/\partial
\mathbf{k}}{\partial\mathcal{L}_{0}/\partial
\omega}=\frac{\partial\omega}{\partial\mathbf{k}},
\end{equation}
and thus we can put $\tilde{\psi}^{(2)}=0$. Next, we introduce a
coordinate system moving with group velocity $\mathbf{v}_{g}$, so
that
\begin{equation}
\label{dT} \frac{\partial}{\partial
T}=-\mathbf{v}_{g}\cdot\frac{\partial}{\partial\mathbf{X}},
\end{equation}
and the spatial variable $\mathbf{X}$ is replaced by
$\mathbf{X}^{\prime}=\mathbf{X}-\mathbf{v}_{g}T$ (and the prime
will be further omitted).

In the $O(\varepsilon^{3})$, one can obtain
\begin{gather}
\mathcal{L}_{0}\tilde{\psi}^{(3)}=-\mathcal{L}_{2}\tilde{\psi}^{(1)}+\frac{\partial}{\partial
x}\left(\frac{\partial \tilde{\psi}^{(1)}}{\partial
x}\frac{\partial \bar{\chi}}{\partial Z}-\frac{\partial
\tilde{\psi}^{(1)}}{\partial z}\frac{\partial \bar{\chi}}{\partial
X}\right) \nonumber \\
 +\left(\frac{\partial \bar{\psi} }{\partial
X}\frac{\partial \tilde{\chi}^{(1)}}{\partial z}-\frac{\partial
\bar{\psi}}{\partial Z}\frac{\partial \tilde{\chi}^{(1)}}{\partial
x}\right) -\left(\frac{\partial \bar{\psi} }{\partial
X}\frac{\partial \Delta\tilde{\psi}^{(1)}}{\partial
z}-\frac{\partial \bar{\psi}}{\partial Z}\frac{\partial
\Delta\tilde{\psi}^{(1)}}{\partial x}\right) , \label{third-order}
\end{gather}
where
\begin{equation}
\label{L2}
\mathcal{L}_{2}=\frac{\partial\mathcal{L}_{0}}{\partial\omega}\left(i\frac{\partial}{\partial
\tau}+\frac{1}{2}\frac{\partial^{2}\omega}{\partial
k_{x}^{2}}\frac{\partial^{2}}{\partial X^{2}}
+\frac{1}{2}\frac{\partial^{2}\omega}{\partial
k_{z}^{2}}\frac{\partial^{2}}{\partial Z^{2}} +
\frac{\partial^{2}\omega}{\partial k_{x}\partial
k_{z}}\frac{\partial^{2}}{\partial X\partial Z}\right),
\end{equation}
and $\partial\mathcal{L}_{0}/\partial\omega=2\omega
(k^{2}+1/4H^{2})$. Removing the secular terms, that is, equating
to zero the right hand side of Eq. (\ref{third-order}), and using
Eqs. (\ref{Psi}) and (\ref{Phi}), we have
\begin{gather}
 \mathcal{L}_{2}\Psi+k_{x}\Psi
\left(k_{x}\frac{\partial\bar{\chi}}{\partial
Z}-k_{z}\frac{\partial\bar{\chi}}{\partial X}\right)+k_{x}\Phi
\left(k_{z}\frac{\partial\bar{\psi}}{\partial
X}-k_{x}\frac{\partial\bar{\psi}}{\partial Z}\right) \nonumber
\\
+ \omega k^{2}\Psi\left(k_{x}\frac{\partial\bar{\psi}}{\partial
Z}-k_{z}\frac{\partial\bar{\psi}}{\partial X}\right)=0,
\label{third-order1}
\end{gather}
To get further progress, we use the linear response for $\Phi$ and
$\bar{\chi}$ from equation (\ref{main2}),
\begin{equation}
\label{linear-responce}
\Phi=-\frac{k_{x}\omega_{g}^{2}}{\omega}\Psi, \quad \frac{\partial
\bar{\chi}}{\partial T}-\omega_{g}^{2}\frac{\partial
\bar{\psi}}{\partial X}=0.
\end{equation}
Next, in the second equation we use Eq. (\ref{dT}). As noted
above, in the following we will be interested in obtaining a
one-dimensional NLS equation for the envelope $\Psi$ containing
either $X$ or $Z$ space variables. Then, from Eq.
(\ref{linear-responce}) we have
$\bar{\chi}=-\omega_{g}^{2}\bar{\psi}/v_{gx}$, where
$v_{gx}=\partial\omega/\partial k_{x}$, if $\partial/\partial
Z=0$, and $\bar{\chi}=0$ if $\partial/\partial X=0$. Thus, Eq.
(\ref{third-order1}) becomes
\begin{equation}
\label{third-order2}
\mathcal{L}_{2}\Psi+k_{z}\left(k_{x}\frac{\omega_{g}^{2}}{v_{gx}}-\omega
k^{2}-\frac{k_{x}^{2}\omega_{g}^{2}}{\omega}\right)\Psi\frac{\partial\bar{\psi}}{\partial
X}+k_{x}\left(\omega k^{2}
+\frac{k_{x}^{2}\omega_{g}^{2}}{\omega}\right)\Psi\frac{\partial\bar{\psi}}{\partial
Z}=0 .
\end{equation}
In the order $O(\varepsilon^{3})$ from (\ref{main3}) for the mean
flow we have the equation
\begin{equation}
\label{LF} \omega_{g}^{2}\frac{\partial^{2}\bar{\psi}}{\partial
X^{2}}-\frac{1}{4H^{2}}\frac{\partial^{2}\bar{\psi}}{\partial
T^{2}}=\overline{\partial_{x}\{\psi,\chi\}}-\overline{\partial_{t}\{\psi,\Delta\psi\}},
\end{equation}
where the bar means averaging over the fast variables. From Eq.
(\ref{LF}), using Eqs. (\ref{dT}) and (\ref{linear-responce}), we
get
\begin{equation}
\label{LF1} \omega_{g}^{2}\frac{\partial^{2}\bar{\psi}}{\partial
X^{2}}-\frac{1}{4H^{2}}\left(v_{gx}^{2}\frac{\partial^{2}\bar{\psi}}{\partial
X^{2}}+v_{gz}^{2}\frac{\partial^{2}\bar{\psi}}{\partial
Z^{2}}\right) = \left(\omega
k^{2}+\frac{k_{x}^{2}\omega_{g}^{2}}{\omega}\right)\left(k_{z}\frac{\partial
|\Psi|^{2}}{\partial X}-k_{x}\frac{\partial |\Psi|^{2}}{\partial
Z}\right),
\end{equation}
where for $v_{gx}$ and $v_{gz}=\partial\omega/\partial k_{z}$ we
have
\begin{equation}
\label{vgx-vgz}
v_{gx}=\frac{\omega_{g}(k_{z}^{2}+1/4H^{2})}{(k^{2}+1/4H^{2})^{3/2}},
\quad v_{gz}=-\frac{\omega_{g}k_{x}k_{z}}{(k^{2}+1/4H^{2})^{3/2}}.
\end{equation}
Equations (\ref{third-order2}) and (\ref{LF1}) are a closed system
of nonlinear equations for the envelope $\Psi$ and the mean flow
$\bar{\psi}$.

\begin{figure}
\centering
\includegraphics[width=3.5in]{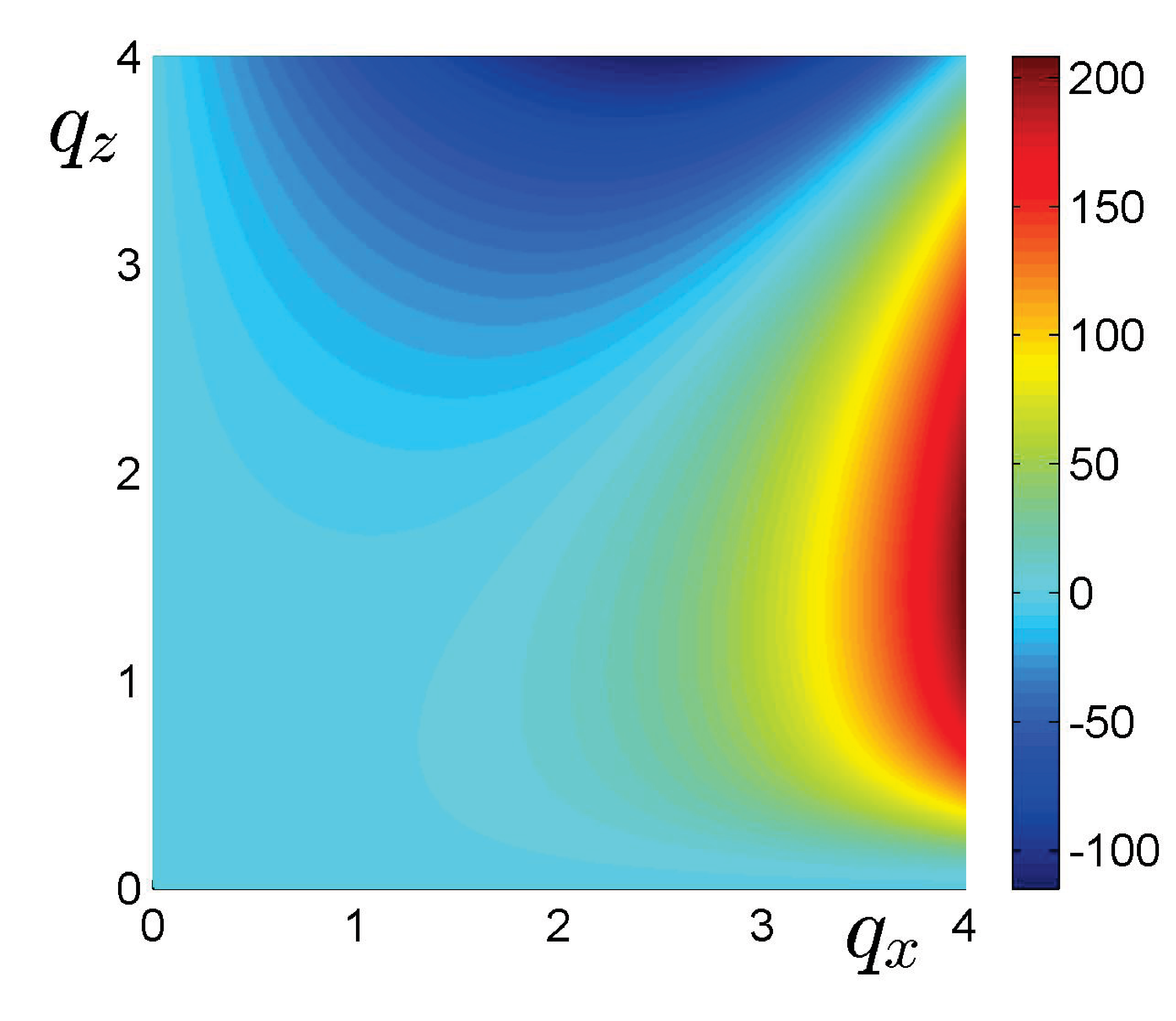}
\caption{\label{fig1} The contour plot of the function
$Q(q_{x},q_{z})$. Negative $Q$ corresponds to the focusing NLS
equation (\ref{NLS}) and positive to the defocusing one.}
\end{figure}

\begin{figure}
\centering
\includegraphics[width=3.5in]{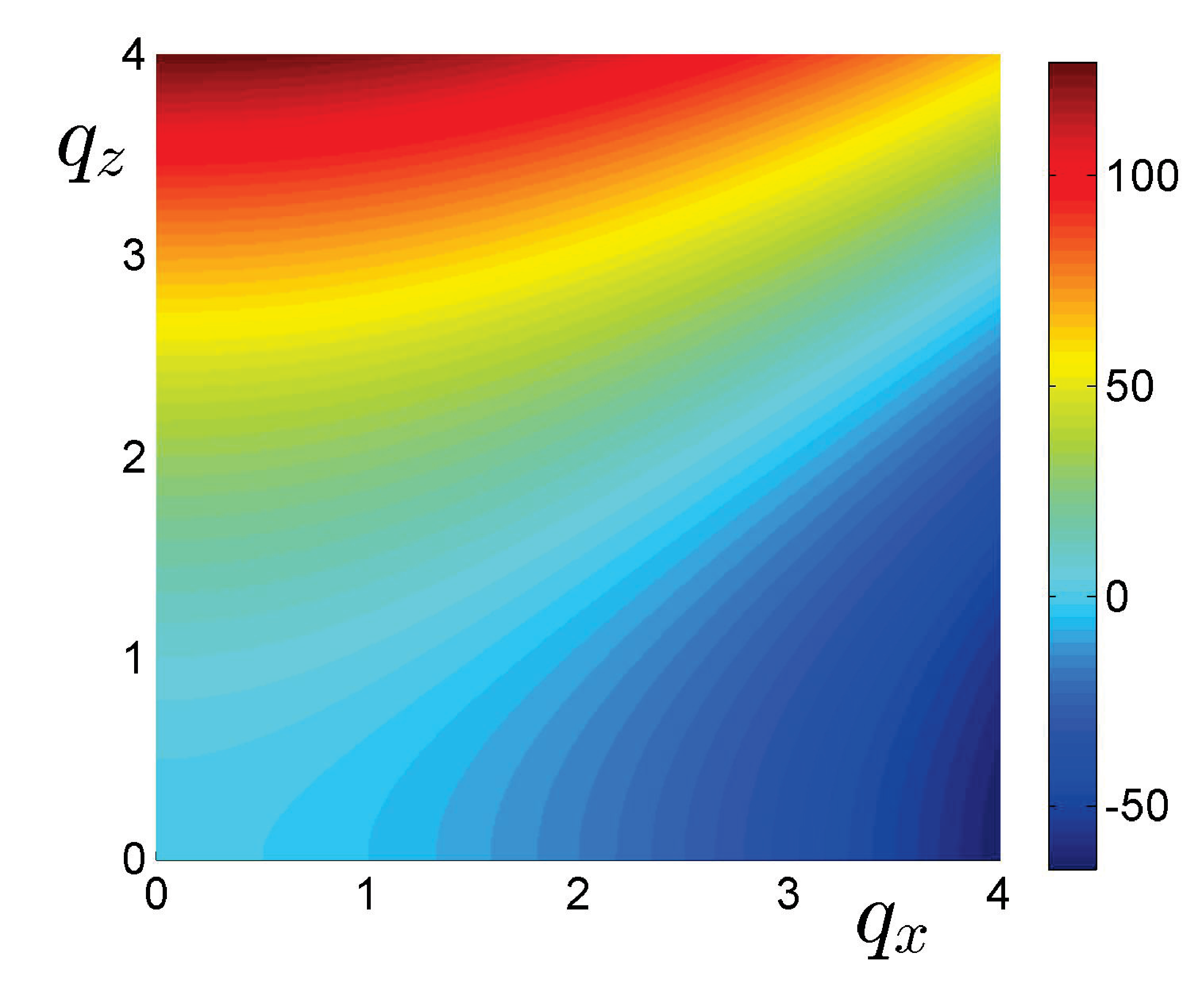}
\caption{\label{fig2} The contour plot of the function
$f(q_{x},q_{z})=8q_{z}^{2}-4q_{x}^{2}-1$ in Eq. (\ref{P1}).
Positive $f$ corresponds to the focusing NLS equation (\ref{NLS1})
and negative to the defocusing one.}
\end{figure}

\section{\label {Sec4} Derivation of nonlinear Schr\"{o}dinger equation}

In this section, we obtain a one-dimensional NLS equation for two
cases. In the first case, the spatial dependence corresponds to
the $X$ coordinate (horizontal propagation), and in the second -
to the $Z$ coordinate (vertical propagation).

Neglecting in Eq. (\ref{third-order2}) the dependence on the
spatial coordinate $Z$, we have
\begin{equation}
\label{NLS-X} 2\omega
\left(k^{2}+\frac{1}{4H^{2}}\right)\left(i\frac{\partial\Psi}{\partial\tau}
+\frac{1}{2}\frac{\partial^{2}\omega}{\partial
k_{x}^{2}}\frac{\partial^{2}\Psi}{\partial X^{2}}\right)
+k_{z}\left(k_{x}\frac{\omega_{g}^{2}}{v_{gx}}-\omega
k^{2}-\frac{k_{x}^{2}\omega_{g}^{2}}{\omega}\right)\Psi\frac{\partial\bar{\psi}}{\partial
X}=0,
\end{equation}
where
\begin{equation}
\frac{\partial^{2}\omega}{\partial
k_{x}^{2}}=-\frac{3k_{x}\omega_{g}(k_{z}^{2}+1/4H^{2})}{(k_{x}^{2}+k_{z}^{2}+1/4H^{2})^{5/2}}.
\end{equation}
Omitting $Z$-dependence in Eq. (\ref{LF1}), one can obtain
\begin{equation}
\label{mean-X} \frac{\partial\bar{\psi}}{\partial
X}=\frac{k_{z}(\omega
k^{2}+k_{x}^{2}\omega_{g}^{2}/\omega)}{\omega_{g}^{2}-v_{gx}^{2}/(4H^{2})}|\Psi|^{2}.
\end{equation}
Next, we introduce dimensionless variables $\tau^{\prime}$,
$X^{\prime}$, and $\Psi^{\prime}$ (recall that $\Psi$ is the
envelope velocity stream function) by
\begin{equation}
\label{dimensionless} \tau^{\prime}=\omega_{g}\tau , \quad
X^{\prime}=\frac{X}{H} , \quad
\Psi^{\prime}=\frac{\Psi}{\omega_{g}H^{2}} ,
\end{equation}
and further the primes are omitted. Then, substituting Eq.
(\ref{mean-X}) into Eq. (\ref{NLS-X}), and taking into account the
explicit expressions (\ref{dispers}) and (\ref{vgx-vgz}) for
$\omega$ and $v_{gx}$ respectively, we have the NLS equation,
\begin{equation}
\label{NLS}
i\frac{\partial\Psi}{\partial\tau}+P\frac{\partial^{2}\Psi}{\partial
X^{2}}+Q|\Psi|^{2}\Psi=0,
\end{equation}
where the dimensionless coefficients $P$ and $Q$ are defined as
\begin{equation}
\label{P} P=-\frac{12q_{x}(1+4q_{z}^{2})}{(1+4q^{2})^{5/2}},
\end{equation}
and
\begin{equation}
\label{Q}
Q=\frac{q_{x}q_{z}^{2}(1+8q^{2})(4q_{x}^{4}-4q_{z}^{4}-q_{z}^{2})}
{(1+4q_{z}^{2})(1+4q^{2})^{3/2}[1-(1+4q_{z}^{2})^{2}/(1+4q^{2})^{3}]},
\end{equation}
respectively, with $q_{x}=k_{x}H$, $q_{z}=k_{z}H$ and
$q^{2}=q_{x}^{2}+q_{z}^{2}$. Note that depending on $k_{x}$ and
$k_{x}$, that is, on the horizontal and vertical wavelengths, and
the effective height of the atmosphere $H$, the $Q$ value can be
either positive or negative. The contour plot of the function
$Q(q_{x},q_{z})$ on the $q_{x}-q_{z}$ plane is shown in
Fig.~\ref{fig1}.

For the vertical propagation, neglecting in Eq.
(\ref{third-order2}) the dependence on the spatial coordinate $X$,
we have
\begin{equation}
2\omega
\left(k^{2}+\frac{1}{4H^{2}}\right)\left(i\frac{\partial\Psi}{\partial\tau}
+\frac{1}{2}\frac{\partial^{2}\omega}{\partial
k_{z}^{2}}\frac{\partial^{2}\Psi}{\partial Z^{2}}\right)
+k_{x}\left(\omega k^{2}
+\frac{k_{x}^{2}\omega_{g}^{2}}{\omega}\right)\Psi\frac{\partial\bar{\psi}}{\partial
Z}=0,
\end{equation}
where
\begin{equation}
\frac{\partial^{2}\omega}{\partial
k_{z}^{2}}=\frac{k_{x}\omega_{g}(2k_{z}^{2}-k_{x}^{2}-1/4H^{2})}{(k^{2}+1/4H^{2})^{5/2}}.
\end{equation}
Then, from Eq. (\ref{LF1}) one obtains
\begin{equation}
\label{mean-Z} \frac{\partial\bar{\psi}}{\partial
Z}=\frac{4k_{x}H^{2}(\omega
k^{2}+k_{x}^{2}\omega_{g}^{2}/\omega)}{v_{gz}^{2}}|\Psi|^{2}.
\end{equation}
Introducing the dimensionless variable $Z^{\prime}=Z/H$ (then the
prime is omitted), and using the dimensionless variables Eq.
(\ref{dimensionless}) for $\tau$ and $\Psi$, we obtain NLS
equation
\begin{equation}
\label{NLS1}
i\frac{\partial\Psi}{\partial\tau}+\tilde{P}\frac{\partial^{2}\Psi}{\partial
Z^{2}}+\tilde{Q}|\Psi|^{2}\Psi=0,
\end{equation}
where the dimensionless coefficients $\tilde{P}$ and $\tilde{Q}$
are defined as
\begin{equation}
\label{P1}
\tilde{P}=\frac{4q_{x}(8q_{z}^{2}-4q_{x}^{2}-1)}{(1+4q^{2})^{3}},
\end{equation}
and
\begin{equation}
\label{Q1}
\tilde{Q}=\frac{q_{x}(1+8q^{2})(1+4q^{2})^{3/2}}{64q_{z}^{2}},
\end{equation}
respectively. The $\tilde{P}$ coefficient at the dispersion term
of Eq. (\ref{NLS1}) has an indefinite sign. The contour plot of
the function $f(q_{x},q_{z})=8q_{z}^{2}-4q_{x}^{2}-1$ is presented
in Fig.~\ref{fig2}.

\section{\label {Sec5} Breathers and rogue waves}

If the coefficients at the dispersion and nonlinear terms in Eqs.
(\ref{NLS}) and (\ref{NLS1}) have the same signs, so that $PQ>0$
and $\tilde{P}\tilde{Q}>0$ , then the corresponding equations have
the form of a focusing NLS equation, otherwise, if $PQ<0$ and
$\tilde{P}\tilde{Q}<0$, the NLS equation has a defocusing type. In
what follows, for definiteness, we will consider for the time
being Eq. (\ref{NLS}) (horizontal propagation). The exact solution
of Eq. (\ref{NLS}) in the form of a plane wave with the frequency
depending on the amplitude $A$ is
\begin{equation}
\label{plane} \Psi=A\mathrm{e}^{iQA^{2}\tau}.
\end{equation}
The standard linear stability analysis then shows that a linear
modulation with the frequency $\Omega$ and the wave number
$\kappa$ obeys the dispersion relation
\begin{equation}
\label{plane-dispers} \Omega^{2}=P\kappa^{2}(P\kappa^{2}-2QA^{2}),
\end{equation}
whose right-hand side is positive if $PQ<0$ and then $\Omega$ is
real. In this case, the modulations of the plane wave are stable,
and for nonvanishing boundary conditions the defocusing NLS
equation has solutions in the form of the so-called dark solitons
\cite{Faddeev1987,Akhmediev1997}. Otherwise, if $PQ>0$, the plane
wave turns out to be unstable with respect to modulations
$\kappa<\sqrt{2Q/P}A$, and for boundary conditions falling off at
infinity, this corresponds, in particular, to bright solitons.
Other solutions of the NLS equation (both focusing and defocusing)
include nonlinear periodic cnoidal waves that can be expressed in
terms of the Jacobi elliptic functions and theta functions
\cite{Akhmediev1997,Chow1995}. Since in our case $P<0$, the type
of the NLS equation (\ref{NLS}) depends on the sign of $Q$. In
this section we consider the case $Q<0$.

Then, the nonlinear stage of the modulation instability
\cite{Zakharov2013} of a plane wave Eq. (\ref{plane})  (also known
as a Benjamin-Feir instability) results in the so-called Akhmediev
breather
\cite{Akhmediev1997,Ahmediev1985,Ahmediev1986,Ahmediev1987},
\begin{equation}
\label{Ahmediev} \Psi_{A}
(X,\tau)=A\mathrm{e}^{iQA^{2}\tau}\left[\frac{\nu^{2}\cosh
(QA^{2}\sigma\tau)+i\sigma\sinh (QA^{2}\sigma\tau)}{\cosh
(QA^{2}\sigma\tau)-\sqrt{1-\nu^{2}/2}\cos (KX)}-1\right],
\end{equation}
where $A$ is the background amplitude,
$\sigma=\nu\sqrt{2-\nu^{2}}$ is the modulation growth rate and
$K=\nu\sqrt{QA^{2}/P}$ is the amplitude-dependent wave number of
the envelope. This solution is localized in the temporal variable
$\tau$ and is periodic in the spatial variable $X$. Both $A$ and
$\nu$ are free real parameters. The solution exists only if
$\nu^{2}<2$, that is, if $K<\sqrt{2Q/P}A$, which is fully
consistent with the modulation instability condition of the plane
wave presented below. Thus, Akhmediev  breather can be treated as
a non-stationary soliton excitation against a constant background
(plane wave). This excitation results in the amplification of the
background wave amplitude. The maximum amplitude takes place at
$t=0$ and in the locations given by the condition $\cos (KX)=1$.
At these locations, the so called amplification factor (ratio of
maximum amplitude to background) is given by
\begin{equation}
\label{amplification}
F_{A}=\frac{\nu^{2}}{1-\sqrt{1-\nu^{2}/2}}-1.
\end{equation}
For $0<\nu^{2}<2$, the amplification factor $F_{A}$ ranges from
$1$ to $3$. The wave energy flux is defined as
\begin{equation}
\label{flux1} J=iP\left(\Psi\frac{\partial\Psi^{\ast}}{\partial
X}-\Psi^{\ast}\frac{\partial\Psi}{\partial X}\right)
\end{equation}
and is related to the energy density $|\Psi|^{2}$ by the
conservation law
\begin{equation}
\label{conservat-low} \frac{\partial |\Psi|^{2}}{\partial
\tau}+\frac{\partial J}{\partial X}=0.
\end{equation}
\begin{figure}
\centering
\includegraphics[width=4.8in]{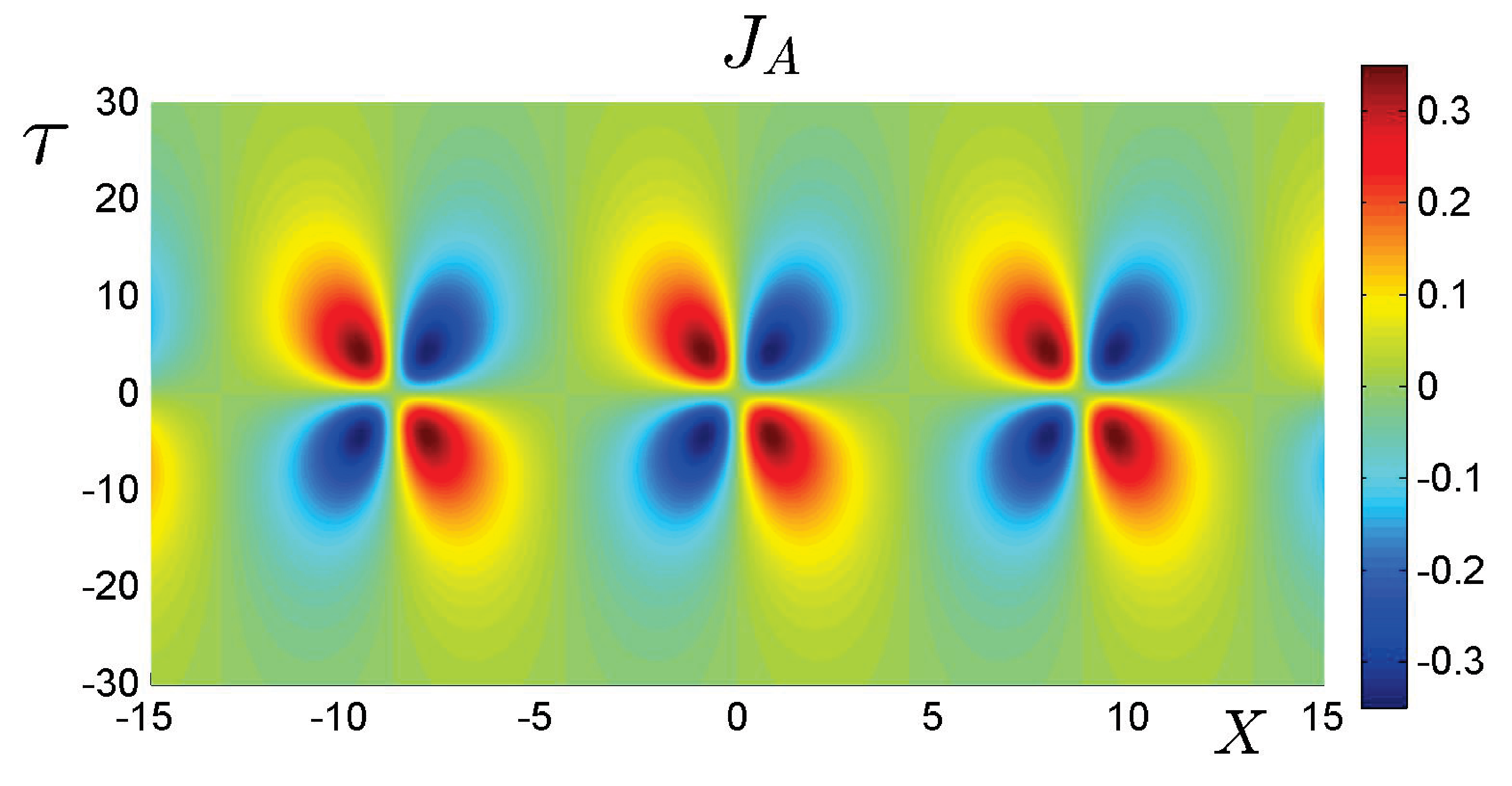}
\caption{\label{fig3} The contour plot of the wave energy flux
$J_{A}(X,\tau)$ in the $X-\tau$ plane for the Akhmediev breather
and the parameters $A=1$, $\nu=1$, $q_{x}=1$ and $q_{z}=3$. It can
be seen that the flux is strongly localized in time (burst of
energy) and at the same time there is a periodic redistribution of
energy in space.}
\end{figure}
For the Akhmediev breather we have,
\begin{equation}
\label{flux-A}
J_{A}(X,\tau)=\frac{\sqrt{2}PA^{2}K\nu(2-\nu^{2})\sin (KX)\sinh
(w\tau)}{[\cosh(w\tau)-\sqrt{1-\nu^{2}/2}\cos(KX)]^{2}},
\end{equation}
where $w=|Q|A^{2}\sigma$. The wave energy flux of the Akhmediev
breather is exponentially localized in time and represents a burst
of energy with a characteristic duration $\tau_{A}\sim 1/w$. At
the same time, a periodic redistribution of energy occurs in space
with a spatial period $X_{A}\sim 2\pi/K$. In a part of space, the
energy flux $J_{A}$ is positive due to the modulation instability
of the background, and in another part, the flux is negative due
to the nonlinear stabilization of the instability. The contour
plot of the wave energy flux (\ref{flux-A}) with $A=1$, $\nu=1$,
$q_{x}=1$ and $q_{z}=3$ (herewith $P=-0.04$ and $Q=-24.7$) is
shown in Fig.~\ref{fig3}. For the Earth's atmosphere, the
effective height of the atmosphere at the considered altitudes
$\gtrsim 200$ km (i.e. for an isothermal atmosphere) is $H\sim 40$
km, i.e. the values of $q_{x}$ and $q_{z}$ correspond to
horizontal and vertical wavelengths  $\sim 40$ km and $\sim 13$
km, respectively.

Another non-stationary solution of Eq. (\ref{NLS}) with $Q<0$ is
known as the so-called Kuznetsov-Ma breather (often referred to as
the Ma breather), which is periodic in time and localized in space
variable,
\begin{equation}
\label{Ma} \Psi_{M}
(X,\tau)=A\mathrm{e}^{iQA^{2}\tau}\left[1+\frac{\mu^{2}\cos
(QA^{2}\rho\tau)+i\rho\sin (QA^{2}\rho\tau)}{\cos
(QA^{2}\rho\tau)-\sqrt{1+\mu^{2}/2}\cosh (\tilde{K}X)}\right],
\end{equation}
where $A$ is the amplitude, $\rho=\mu\sqrt{2+\mu^{2}}$ and
$\tilde{K}=\mu\sqrt{QA^{2}/P}$. Free real parameters in (\ref{Ma})
are $A$ and $\mu$. This solution was first found by Kuznetsov
using the inverse scattering transform method
\cite{Kuznetsov1977}, and was rediscovered in
\cite{Kawata1978,Ma1979} as well as others later on. The period of
oscillations in the Kuznetsov-Ma breather is
$T=2\pi/(|Q|A^{2}\mu\sqrt{2+\mu^{2}})$.
\begin{equation}
\label{flux-M}
J_{M}(X,\tau)=\frac{\sqrt{2}PA^{2}\tilde{K}\mu(2+\mu^{2})\sinh
(\tilde{K}X)\sin
(w\tau)}{[\cos(w\tau)-\sqrt{1+\mu^{2}/2}\cosh(\tilde{K}X)]^{2}},
\end{equation}
where $\omega=|Q|A^{2}\rho$.
\begin{figure}
\centering
\includegraphics[width=4.8in]{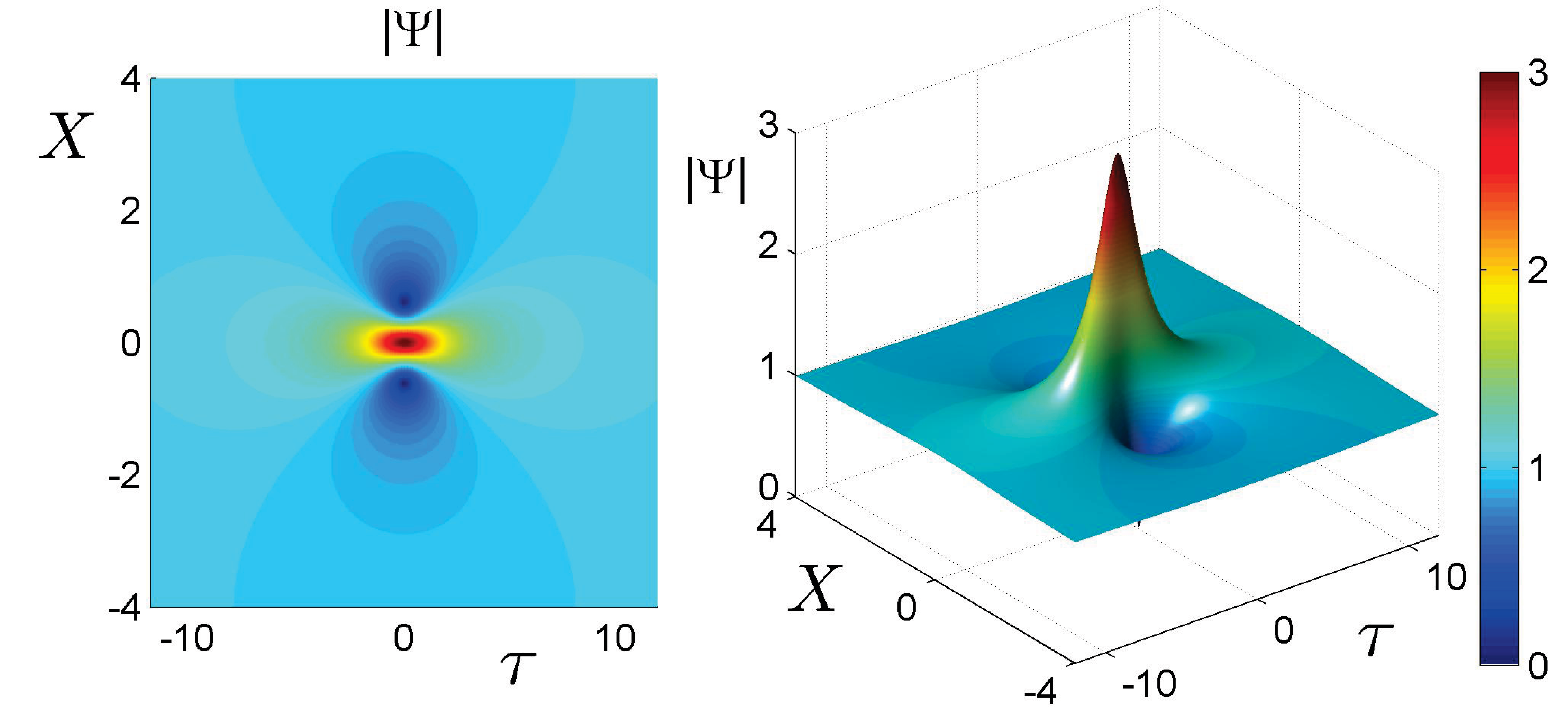}
\caption{\label{fig4} The rogue wave (\ref{Peregrin}) with $A=1$,
$q_{x}=1.5$, $q_{z}=1.5$. Left column: two-dimensional contour
plot of the amplitude $|\Psi(X,\tau)|$ in the $X-\tau$ plane;
right column: three-dimensional view. It can be seen that the wave
appearing from nowhere and disappearing without a trace.}
\end{figure}
The wave energy flux for the Kuznetsov-Ma breather is localized in
space with a characteristic size $X_{M} \sim 1/\tilde{K}$ and
represents periodic bursts of wave energy with a period
$\tau_{M}\sim 2\pi/w$.

The limiting case of both the Akhmediev breather (\ref{Ahmediev})
with $\nu\rightarrow 0$, and the Kuznetsov-Ma breather (\ref{Ma})
with $\mu\rightarrow 0$ corresponds to the Peregrine soliton
\cite{Peregrine1983}
\begin{equation}
\label{Peregrin} \Psi_{P}
(X,\tau)=A\mathrm{e}^{iQA^{2}\tau}\left[\frac{4(1+2iQA^{2}\tau)}{1+4Q^{2}A^{4}\tau^{2}
+2QA^{2}X^{2}/P}-1\right].
\end{equation}
In fact, this non-stationary solution, strictly speaking, is not a
soliton, but is a rogue wave that appears against a constant
background from nowhere and disappears without a trace. The
characteristic lifetime of the rogue wave can be estimated as
$\tau_{c}\sim 1/(2Q^{2}A)$. The amplification factor for the
Peregrine soliton is $F_{P}=3$. Despite some idealization of the
model, the solution in the form of a Peregrin soliton agrees very
well with observations and experimental data. For example,
numerous observations, starting from the very first on rogue waves
in the ocean (including the first famous sighting on the Draupner
platform in the North Sea off the coast of Norway on 1 January
1995) \cite{Dysthe2008,Pelinovski2009}, show an amplification
factor  $\sim 3$. The rogue solution (\ref{Peregrin}) with $A=1$,
$q_{x}=1.5$ and $q_{z}=1.5$ (herewith $P=-0.11$ and $Q=-0.34$) is
shown in Fig.~\ref{fig4}. Such values of $q_{x}$ and $q_{z}$
correspond to horizontal and vertical wavelengths  $\sim 20$ km.
The peak and two troughs are visible against the background, which
corresponds to the conservation of the norm $N=\int
(|\Psi|^{2}-A^{2})dX$ for the NLS equation (\ref{NLS}).

For the vertical propagation corresponding to equation
(\ref{NLS1}), the breather solutions and the rogue solution are
obtained from Eqs. (\ref{Ahmediev}), (\ref{Ma}) and Eq.
(\ref{Peregrin}) by replacing $X\rightarrow Z$, and
$P\rightarrow\tilde{P}$, $Q\rightarrow\tilde{Q}$.

\section{\label {Sec6} Periodic nonlinear waves and dark solitons}

In this section, we will consider the case when $PQ<0$ and
$\tilde{P}\tilde{Q}<0$ in Eqs. (\ref{NLS}) and (\ref{NLS1})
respectively. As follows from Eqs. (\ref{P1}) and (\ref{Q1}), this
corresponds, in particular, to sufficiently long vertical
wavelengths compared to the effective height $H$ and horizontal
wavelengths. For example, for vertical propagation, the condition
$k_{z}^{2}<k_{x}^{2}/2+1/8H^{2}$ must be met. As in the previous
section, for definiteness, we consider the case of horizontal
propagation, i.e. equation (\ref{NLS}). The case of vertical
propagation is obtained in the following equations by replacing
$X\rightarrow Z$, and $P\rightarrow\tilde{P}$,
$Q\rightarrow\tilde{Q}$.

The defocusing NLS equation (\ref{NLS}) with $PQ<0$ has a
stationary solution in the form of a nonlinear periodic (cnoidal)
wave \cite{Akhmediev1997},
\begin{equation}
\label{knoidal} \Psi
(X,\tau)=\frac{\sqrt{2}A\mathrm{e}^{iQA^{2}\tau}m}{\sqrt{1+m^{2}}}
\mathrm{sn}\left( A\sqrt{\frac{Q}{P(1+m^{2})}}X,m\right),
\end{equation}
where $A$ is the free real parameter, $\mathrm{sn}(u,m)$ is the
Jacobi elliptic sine with the modulus $m$. In the particular case
$m=1$, solution (\ref{knoidal}) takes the form of a dark soliton,
\begin{equation}
\label{dark1} \Psi (X,\tau)=A\mathrm{e}^{iQA^{2}\tau}\tanh
\left(A\sqrt{\frac{Q}{2P}}X\right).
\end{equation}
Since the NLS equation is Galilean invariant, solutions moving
with the velocity $V$ can be obtained from solutions at rest
(\ref{knoidal}) and (\ref{dark1}) by replacing $X\rightarrow
X-V\tau$ and replacing the exponential factor
\begin{equation}
\exp(iQA^{2}\tau)\rightarrow
\exp\left(iQA^{2}\tau+i\sqrt{\frac{Q}{2P}}\frac{V}{2}X-iQ\frac{V^{2}}{4}\tau\right).
\end{equation}
The dark soliton (\ref{dark1}) has a dip in $A$ against a uniform
background $A$ and is the so-called black soliton. A more general
solution (gray solitons), corresponding to the dip of the
amplitude of smaller $A$ against the background $A$, has the form
\begin{gather}
\Psi
(X,\tau)=A\left\{i\sin\varphi+\cos\varphi\tanh\left[a\sqrt{\frac{Q}{2P}}
(X-V\tau)\right]\right\} \nonumber \\
\times\exp\left[i\sqrt{\frac{Q}{2P}}\frac{c}{2}X+iQ\left(A^{2}-\frac{c^{2}}{4}\right)\tau\right],
\label{dark2}
\end{gather}
where $c$ is the free real parameter, $\tan\varphi=(c-V)/a$, and
the soliton parameters are connected by the constraint
$a^{2}+(c-V)^{2}=A^{2}$, otherwise written as $a=\pm
A\cos\varphi$. Thus, the soliton is characterized by three
independent parameters $A$, $V$ (or $c$), and $\varphi$. The
soliton therefore exists in the domain $-A+c<V<A+c$.
\begin{figure}
\centering
\includegraphics[width=4.8in]{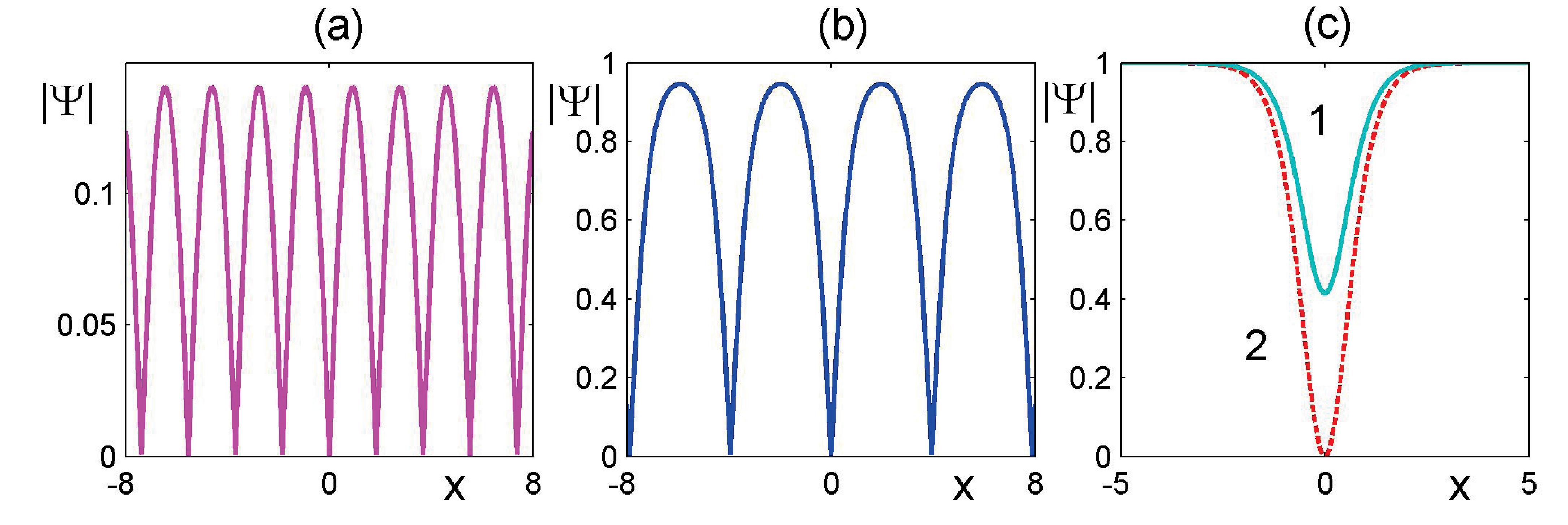}
\caption{\label{fig5} a) Nonlinear periodic wave (\ref{knoidal})
with $A=1$, $q_{x}=1$, $q_{z}=0.5$ and the modulus $m=0.1$ ; b)
the same with the modulus $m=0.9$; c) dark solitons (\ref{dark2}):
the solid line 1 - $\varphi=0.7$, gray soliton; the dotted line 2
- $\varphi=0$, black soliton.}
\end{figure}
It describes a localized kink structure moving with the velocity
$V$ on a background plane wave; $A$ gives the amplitude of the
background. Across the soliton, there is a phase jump in the
background wave of $\pi-2\varphi$. For the intensity $|\Psi|^{2}$
we have
\begin{equation}
\label{dark-amplitude} |\Psi
(X,\tau)|^{2}=A^{2}\left\{1-\cos^{2}\varphi\,\mathrm{sech}^{2}
\left[a\sqrt{\frac{Q}{2P}}(X-V\tau)\right]\right\}.
\end{equation}
The contrast of the dark soliton, defined as the ratio between the
maximum and minimum intensities, is given by $\cos^{2}\varphi$.
For $\varphi=0$ we have the black soliton. The nonlinear periodic
waves (\ref{knoidal}) with the modulus $m=0.1$ and $m=0.9$, and
the dark solitons with $\varphi=0.7$ (gray soliton) and
$\varphi=0$ (black soliton) are presented in Fig.~\ref{fig5}.
Other parameters are $A=1$, $q_{x}=1$ and $q_{z}=0.5$ (herewith
$P=-0.27$ and $Q=0.33$). For the Earth's atmosphere, such values
of $q_{x}$ and $q_{z}$ correspond to horizontal and vertical
wavelengths  $\sim 40$ km and $\sim 80$ km, respectively.

\section{\label {Sec7} Conclusion}

In this work, we have applied the reductive perturbation method to
obtain model nonlinear equations describing the dynamics of IGWs
in the atmosphere. A system of two-dimensional nonlinear equations
for the velocity stream function and the mean flow has been
derived in the envelope approximation. In the one-dimensional
case, we have obtained the NLS equation for the envelope
corresponding to both horizontal and vertical propagation of IGWs.
Depending on the values of the horizontal and vertical
wavelengths, this equation can be either focusing (the signs of
the dispersion and nonlinear terms are the same) or defocusing. In
the focusing case, non-stationary solutions in the form of the
Peregrine soliton (rogue wave), the Akhmediev breather and the
Kuznetsov-Ma breather have been considered as potential candidates
for the modeling of rogue waves in the atmosphere. In the
defocusing case, stationary nonlinear IGWs have been found in the
form of nonlinear periodic waves and dark solitons.

We have considered the approximation of an isothermal atmosphere,
which, for the Earth's atmosphere, in particular, is fully
justified at altitudes $\gtrsim 200$ km. Note, however, that other
altitude intervals can be distinguished in the Earth's atmosphere,
where the temperature changes so slowly that its change can be
neglected within sufficiently thin layers (the so-called
isothermal layers). The propagation of waves at such altitudes can
also be described in terms of the theory of an isothermal
atmosphere.

Note that in this paper we have restricted ourselves to
one-dimensional nonlinear structures in the framework of the
one-dimensional NLS equation. An analysis of the dynamics and the
possibility of the existence of nonlinear two-dimensional
structures within the framework of the system of equations
(\ref{third-order2}) and (\ref{LF1}) will be addressed in a future
work.

\section*{Declaration of competing interest}
The authors declare that they have no known competing financial
interests or personal relationships that could have appeared to
influence the work reported in this paper.

\section*{CRediT authorship contribution statement}
\textbf{V. M. Lashkin}: Conceptualization, Methodology,
Validation, Formal analysis, Investigation. \textbf{O. K.
Cheremnykh}: Conceptualization, Methodology, Validation, Formal
analysis, Investigation.

\section*{Data availability}
No data was used for the research described in the article.

\section*{Acknowledgments}

The work was supported by the National Research Foundation of
Ukraine, grant 2020.02/0015.


\end{document}